\documentclass[a4paper,12pt]{article}
\usepackage[pctex32]{graphics}

\begin{document}
\topmargin 0pt
\oddsidemargin 0mm

\newcommand{\mn}{\mu\nu}
\newcommand{\bta}{\beta}
\newcommand{\gmm}{\gamma}
\newcommand{\del}{\delta}
\newcommand{\omg}{\omega}
\newcommand{\sgm}{\sigma}
\newcommand{\lmd}{\lambda}
\newcommand{\tha}{\theta}
\newcommand{\vph}{\varphi}
\newcommand{\Alp}{\Alpha}
\newcommand{\Bta}{\Beta}
\newcommand{\Gmm}{\Gamma}
\newcommand{\Del}{\Delta}
\newcommand{\Omg}{\Omega}
\newcommand{\Sgm}{\Sigma}
\newcommand{\Lmd}{\Lambda}
\newcommand{\Tha}{\Theta}
\newcommand{\half}{\frac{1}{2}}
\newcommand{\rnd}{\partial}
\newcommand{\nab}{\nabla}
\newcommand{\fr}{\frac}

\newcommand{\beq}{\begin{equation}}
\newcommand{\eeq}{\end{equation}}
\newcommand{\beqa}{\begin{eqnarray}}
\newcommand{\eeqa}{\end{eqnarray}}

\begin{titlepage}

\vspace{5mm}
\begin{center}
{\Large \bf Graviton and scalar propagations on AdS$_4$ space in
$f(R)$ gravities } \vspace{12mm}

{\large   Yun Soo Myung \footnote{e-mail
 address: ysmyung@inje.ac.kr}}
 \\
\vspace{10mm} {\em  Institute of Basic Science and School of
Computer Aided Science, Inje University, Gimhae 621-749, Korea }

\end{center}

\vspace{5mm} \centerline{{\bf{Abstract}}}
 \vspace{5mm}
 We investigate  propagations of graviton
and additional scalar on four-dimensional anti de Sitter (AdS$_4$)
space using $f(R)$ gravity models with external sources. It is shown
that there is the van Dam-Veltman-Zakharov (vDVZ) discontinuity in
$f(R)$ gravity models because $f(R)$ gravity implies GR with
additional scalar. This indicates  a difference between general
relativity and $f(R)$ gravity clearly.
\end{titlepage}
\newpage
\renewcommand{\thefootnote}{\arabic{footnote}}
\setcounter{footnote}{0} \setcounter{page}{2}

\section{Introduction}
There has been much interest in the massless limit of the massive
graviton propagator~\cite{Hig,KMP,POR,GNie,DW}. A key issue of this
approach is that van  Dam-Veltman-Zakharov (vDVZ)
discontinuity~\cite{DVZ} is peculiar to Minkowski space, but it does
not arise in (anti) de Sitter space. The vDVZ discontinuity implies
that  the limit of $M^2 \to 0$ does not yield a massless graviton at
the tree level such that the Einstein gravity (general relativity:
GR) is isolated from the massive gravity.  One has usually
introduced the spin-2 Fierz-Pauli mass term with mass squared
$M^2_{\rm FP}$~\cite{FP} for this calculation. If the cosmological
constant (CC, $\Lambda$) was introduced, the limit of $M^2_{\rm
FP}/\Lambda \to 0$ may recover a massless graviton. Another
resolution to the discontinuity is possible to occur even in
Minkowski space, if the Schwarzschild radius of the scattering
objects is taken  to be the second mass scale~\cite{Vain}. However,
these all belong to the linearized (tree) level calculations. If
one-loop graviton vacuum amplitude is considered for a massive
graviton~\cite{DDLS}, the discontinuity appears again. This means
that the apparent absence of the vDVZ discontinuity may be
considered as  an artifact of the tree level approximation. Also,
there is the Boulware-Deser instability which states that at the
non-linearized level, a ghost appears in the massive gravity
theory~\cite{BD}.

On the other hand,  $f(R)$ gravities  as modified gravity theories
~\cite{NO,sf,NOuh} have much attentions as one of strong candidates
for explaining the current accelerating universe~\cite{SN}.
Actually, $f(R)$ gravities can be considered as GR (massless
graviton) with an additional scalar field.  Explicitly, it was shown
that the metric-$f(R)$ gravity is equivalent to the $\omg_{\rm
BD}=0$ Brans-Dicke (BD) theory with the potential, while the
Palatini-$f(R)$ gravity is equivalent to the $\omg_{\rm BD}=-3/2$ BD
theory with the potential~\cite{FT}.
 However, it was pointed out that the
mapping seems to be problematic because the  potential defined by
$U(\Phi(R))=R \Phi-f(R)$ with $\Phi=\partial_R f(R)$ induces a
singularity in the cosmological evolution~\cite{BENO,BNO,JPS}.
 Although the
equivalence principle test (EPT) in the solar system imposes a
strong constraint on $f(R)$ gravities, they may not be automatically
ruled out if the Chameleon mechanism is introduced to resolve it. It
was shown that the EPT allows  $f(R)$ gravity models that are
indistinguishable from the ${\rm \Lambda}$CDM model ($R$+positive
CC) in the background universe evolution~\cite{PS}. However, this
does not necessarily imply that there is no difference in the
dynamics of perturbations~\cite{CENOZ}.

There were perturbation studies for the propagation of graviton on
the constant curvature background using  a single $f(R)$
gravity~\cite{NS}, but the analysis is not complete because they did
not calculate  one-particle scattering amplitude with  external
sources $T_{\mn}$. Also, it was argued that there is no vDVZ
discontinuity in  GR with higher curvature terms (for example,
$R-2\Lambda +\alpha R^2$) on  AdS$_4$ space~\cite{Neu}. Recently, a
similar analysis was performed in $D$-dimensional anti-de Sitter
(AdS$_{\rm D}$) space, including the new massive gravity in three
dimensions~\cite{GT}.

 In this work, we investigate  propagations of graviton
and additional scalar on AdS$_4$ space using $f(R)$ gravity models
with external sources. Furthermore, we show that the vDVZ
discontinuity appears in  $f(R)$ gravity models because $f(R)$
gravity means GR with an  additional scalar.

\section{$f(R)$ gravities}

We start from  $f(R)$ gravity  without any matter including a
cosmological constant \beq \label{act-0} I =  \frac{1}{16 \pi G}
\int d^4x \sqrt{- g} \Big \{ R +f(R)\Big \}. \label{Action} \eeq
This splitting form of ``$R+f(R)$" is rather treatable than a single
``$f(R)$" form.  In this work, $f(R)$ gravity means  the former form
and we set $16 \pi G=2$. Also we follow the signature of $(-+++)$.
The equation of motion is given by \beq \mbox{}
R_{\mn}\Big[1+f'(R)\Big]-\frac{1}{2}g_{\mn}\Big[R+f(R)\Big] + \Big[
g_{\mn}\nabla^2 -\nabla_\mu \nabla_\nu \Big]f'(R)=0. \label{FEQ}
\eeq For the case of $f(R)=-2\Lambda(f'=f''=0)$ (equivalently, GR
with CC), we have the Einstein-Hilbert action with a negative CC. In
this case, the vacuum solution is
 the four dimensional anti de Sitter (AdS$_4$) space whose geometry is expressed in
terms of the metric ($\bar g_{\mn})$  as \beq \bar
R_{\mu\nu\rho\sigma}=\frac{\Lambda}{3}(\bar g_{\mu\rho}\bar
g_{\nu\sigma}- \bar g_{\mu\sigma}\bar g_{\nu\rho}),~~~ \bar
R_{\mn}=\Lambda \bar g_{\mn},~~~ \bar R=4 \Lambda=-\fr{12}{\ell^2}.
\label{AdS} \eeq Its line element takes the form \beq ds^2_{\rm
AdS}=\bar{g}_{\mn} dx^\mu dx^\nu= -
\Big(1+\fr{r^2}{\ell^2}\Big)dt^2+\frac{dr^2}{\Big(1+\fr{r^2}{\ell^2}\Big)}+r^2d\Omega^2_2.
\eeq
 In order to find a similar AdS$_4$ space solution, one has to
consider a constant curvature scalar  $R=\bar{R}$ with
$f'(\bar{R})={\rm const}$ and $f''(\bar{R})={\rm const}$. In this
case, Eq.(\ref{FEQ}) leads to \beq
\bar{R}_{\mn}\Big[1+f'(\bar{R})\Big]-\frac{1}{2}\bar{g}_{\mn}\Big[\bar{R}+f(\bar{R})\Big]=0
\label{cFEQ} \eeq  which means that the third term in (\ref{FEQ})
plays no role in obtaining the constant curvature solution. However,
it will play an important role in the perturbation analysis around
the background of  AdS$_4$ space. Taking the trace of (\ref{cFEQ}),
$\bar{R}$ is  determined by an algebraic equation \beq
\Big[1+f'(\bar{R})\big]\bar{R}-2\Big[\bar{R}+f(\bar{R})\Big]=0. \eeq
From this equation, on finds the constant curvature scalar  as a
function of $f(\bar{R})$ and $f'(\bar{R})$ \beq \label{bcurv}
\bar{R}=2\Bigg[\frac{\bar{R}+f(\bar{R})}{1+f'(\bar{R})}\Bigg]=\frac{2f(\bar{R})}{f'(\bar{R})-1}\equiv
4\bar{\Lambda}_f, \eeq where the last equivalence is established by
 analogy of (\ref{AdS}). We call $\bar{\Lambda}_f$ an effective
cosmological constant because it is not a genuine CC but it is an
induced CC from $f(R)$ gravities.  Similarly, from (\ref{cFEQ}) we
read off the Ricci tensor \beq
\bar{R}_{\mn}=\frac{1}{2}\Bigg[\frac{\bar{R}+f(\bar{R})}{1+f'(\bar{R})}\Bigg]\bar{g}_{\mn}=\bar{\Lambda}_f\bar{g}_{\mn}.\eeq
Hence, as far as the AdS$_4$ vacuum solution is concerned, there is
no essential difference between GR with CC ($R-2\Lambda$)  and
$f(R)$ gravity.  However, we have to distinguish two models by
noting that $f'(\bar{R})={\rm const}$ and $f''(\bar{R})={\rm const}$
in $f(R)$ gravities.  Furthermore, it is well known  that metric
$f(R)$ gravity (especially for $R+\alpha R^2$~\cite{Whit}) is
equivalent to the $\omega_{BD}=0$ Brans-Dicke theory with the
potential (scalar-tensor theory)\footnote{More explicitly,  the
metric $f(R)$ gravity is equivalent to the Brans-Dicke theory with
the potential  in the Jordan frame, while $f(R)$ gravities of
$R+f(R)$ is equivalent to GR with a scalar field in the Einstein
frame. Hence, the AdS$_4$ space solution is mapped into other
constant curvature solution with specific solution for a scalar
field.}. Therefore we expect from (\ref{act-0}) that a massless
graviton (2 degrees of freedom: 2DOF) and a massive scalar (1 DOF)
propagate on AdS$_4$ space without any ghost.

\section{Perturbation analysis}
In order to study the propagation of the metric, we  introduce the
perturbation around the background  of AdS$_4$ space \beq g_{\mu\nu}
= \bar g_{\mu\nu} +h_{\mu\nu}. \label{Per} \eeq Hereafter we denote
the background values  with ``overbar". After a lengthy calculation,
the linearized equation to Eq.(\ref{FEQ}) with the external source
$T_{\mn}$ takes the form  \beq \label{lineq} (1+f'(\bar{R})) \delta
G_{\mn}(h) +
f''(\bar{R})\Big[\bar{g}_{\mn}\bar{\nabla}^2-\bar{\nabla}_\mu
\bar{\nabla}_\nu+\bar{\Lambda}_f\bar{g}_{\mn}\Big]\delta R(h) =
T_{\mn}, \label{PEQ} \eeq where the linearized Einstein tensor with
an effective CC is given by~\cite{GT}   \beq \label{linEeq}\delta
G_{\mn}(h)=\delta R_{\mn}(h)-\frac{\bar{g}_{\mn}}{2}\delta
R(h)-\bar{\Lambda}_f h_{\mn}. \eeq The linearized Ricci tensor and
the linearized scalar curvature take the forms, respectively,
\begin{eqnarray}
&& \delta R_{\mn}(h)= \frac{1}{2}\Big[\bar{\nabla}^\rho
\bar{\nabla}_{\mu} h_{\nu\rho}+\bar{\nabla}^\rho \bar{\nabla}_{\nu}
h_{\mu\rho}-\bar{\nabla}^2h_{\mn} -\bar{\nabla}_\mu\bar{\nabla}_\nu
h\Big],
\label{eqr} \\
&& \delta R(h)= \bar{g}^{\mn}\delta R_{\mn}(h)-
h^{\mn}\bar{R}_{\mn}= \bar{\nabla}^\rho \bar{\nabla}^{\mu}
h_{\rho\mu}-\bar{\nabla}^2h-\bar{\Lambda}_fh. \label{eqs}
\end{eqnarray}
 In
deriving these, we used the Taylor expansions
around the constant curvature scalar  background $R=\bar{R}$  as \beqa && f(R)= f(\bar{R})+f'(\bar{R})\delta R(h)+\cdots, \\
&& f'(R)= f'(\bar{R})+f''(\bar{R})\delta R(h)+\cdots. \eeqa The
trace of (\ref{lineq}) has \beq \label{tracel} \Big[-(1+f'(\bar{R}))
+f''(\bar{R})\Big(3 \bar{\nabla}^2+4\bar{\Lambda}_f\Big)\Big]\delta
R(h)=T. \eeq
 At
this stage, we note that the linearized equation (\ref{lineq})  is
invariant under linearized diffeomorphisms as \beq \delta_\xi
h_{\mn}=\bar{\nabla}_{\mu} \xi_\nu+ \bar{\nabla}_{\nu} \xi_\mu, \eeq
because of \beq \delta_{\xi} \delta G_{\mn}(h)=0,~~~\delta_\xi
\delta R(h)=0. \eeq This implies that divergence and double
divergence do not provide any constraint on $h_{\mn}$. Also, the
gauge invariant (physical) quantity is still left undetermined by
the linearized equation (\ref{lineq}).

In order to find  physically propagating modes, we decompose the
metric perturbation $h_{\mn}$ with 10 DOF   into \beq
h_{\mn}=h^{TT}_{\mn}+\bar{\nabla}_{(\mu} V_{\nu)}+\bar{\nabla}_{\mu}
\bar{\nabla}_{\nu} \phi +\psi \bar{g}_{\mn}, \eeq where
$h^{TT}_{\mn}$ is the transverse traceless tensor with 5 DOF
($\bar{\nabla}^\mu h^{TT}_{\mn}=0,h^{TT}=0$), $V_\nu$ is a
divergence free vector with 3 DOF ($\bar{\nabla}^\mu V_\mu=0$), and
$\phi$ and $\psi$ are scalar fields with 2 DOF. These imply two
relations \beq \label{relh} \bar{\nabla}^2h=\bar{\nabla}^4\phi+4
\bar{\nabla}^2\psi,~~ \bar{\nabla}^\mu\bar{\nabla}^\nu
h_{\mn}=\bar{\nabla}^4\phi+\bar{\Lambda}_f \bar{\nabla}^2\phi+
\bar{\nabla}^2\psi. \eeq Up to now, we did not make any choice on
the gauge-fixing. One-particle scattering amplitude is mostly
computed by choosing a condition of \beq \label{gcond}
\bar{\nabla}^\mu\bar{\nabla}^\nu h_{\mn}=\bar{\nabla}^2h. \eeq
 When  the  mass term is present,  this
condition could be derived natually~\cite{Neu,GT,Met}.  For example,
adding $M^2_{\rm FP}(h_{\mn}-h\bar{g}_{\mn})/2$ to the linearized
equation (\ref{lineq}) leads to the condition of $\bar{\nabla}^\mu
h_{\mn}=\bar{\nabla}_\nu h$ ((\ref{gcond})) when hitting single
(double) divergence on it~\cite{GT}.  In  $f(R)$ gravity theories,
however,  we do not consider any mass term. Hence, (\ref{gcond})
could be obtained  from a gauge-fixing  of \beq \label{gauge-f}
\bar{\nabla}^\mu h_{\mn}=\bar{\nabla}_\nu h. \eeq Then, considering
(\ref{relh}) together with this condition leads to \beq 3
\bar{\nabla}^2\psi=\bar{\Lambda}_f \bar{\nabla}^2\phi \eeq which
implies that two scalars $\phi$ and $\psi$ are not independent under
the condition of (\ref{gcond}). Plugging this into the first
relation of (\ref{relh}), one finds a relation between the trace of
$h_{\mn}$ and scalar $\psi$ as \beq \label{h-eq}
h=\Big[\frac{3}{\bar{\Lambda}_f}\bar{\nabla}^2+4\Big]\psi. \eeq
Imposing (\ref{gcond}) on (\ref{eqs}) leads to $\delta
R(h)=-\bar{\Lambda}_f h$. Using  (\ref{tracel}) and (\ref{h-eq}), we
express $\psi$ in terms of the trace $T$ of external sources
$T_{\mn}$ as \beq \psi=\frac{
1}{9f''(\bar{R})\Big[\frac{1+f'(\bar{R})}{3f''(\bar{R})}-\Big(\bar{\nabla}^2+
\frac{4}{3}\bar{\Lambda}_f\Big)\Big]\Big(\bar{\nabla}^2+\frac{4}{3}\bar{\Lambda}_f\Big)}T\eeq
which means that  $\psi$  becomes a massive scalar on AdS$_4$ space
of $f(R)$ gravities.

 In order to find  the transverse traceless part $h^{TT}_{\mn}$, we need the
 Lichnerowicz operator $\Delta_L$ acting on spin-2 symmetric
 tensors  defined by
 \beq
 \Delta_L
 h_{\mn}=-\bar{\nabla}^2h_{\mn}+\frac{8\bar{\Lambda}_f}{3}\Bigg(h_{\mn}-\fr{h}{4}\bar{g}_{\mn}\Bigg).
 \eeq
Taking into account this, we rewrite the linearized Einstein tensor
as \beq \delta
G^{TT}_{\mn}(h)=\frac{\Delta_Lh^{TT}_{\mn}}{2}-\bar{\Lambda
}h^{TT}_{\mn}. \eeq Hence, we express $h_{\mn}^{TT}$ in terms of
external sources as \beq
h_{\mn}^{TT}=\frac{2}{(1+f'(\bar{R}))(\Delta_L-2\bar{\Lambda}_f)}T^{TT}_{\mn},
\eeq where the transverse traceless source ($ \nabla^\mu
T_{\mn}^{TT}=0, T^{TT}=0$) is given by~\cite{POR} \beq T_{\mn}^{TT}=
T_{\mn} - \frac{1}{3} T g_{\mn} +\frac{1}{3}(\bar{\nabla}_\mu
\bar{\nabla}_\nu + g_{\mn}\bar{\Lambda}_f/3)(\bar{\nabla}^2 + 4
\bar{\Lambda}_f/3)^{-1}T. \eeq We are now in a position to define
the tree-level (one particle) scattering amplitude between two
external sources $\tilde{T}_{\mn}$ and $T_{\mn}$  as \beq
A=\frac{1}{4} \int d^4x \sqrt{-\bar{g}}\tilde{T}_{\mn}(x) h^{\mn}(x)
\equiv \frac{1}{4}\Big[\tilde{T}_{\mn} h^{TT \mn }+\tilde{T}
\psi\Big], \eeq where we suppress the integral to have a notational
simplicity in the last expression. Finally, the scattering amplitude
takes the form \beqa \label{opamp}4A
&=&2\tilde{T}_{\mn}\Big[(1+f'(\bar{R}))(\Delta_L-2\bar{\Lambda}_f)\Big]^{-1}T^{\mn}+\frac{2}{3}\tilde{T}
\Big[(1+f'(\bar{R}))(\bar{\nabla}^2+2\bar{\Lambda}_f)\Big]^{-1}T
\\ \nonumber
&-&\frac{2\bar{\Lambda}_f}{9}\tilde{T}
\Big[(1+f'(\bar{R}))(\bar{\nabla}^2+2\bar{\Lambda}_f)\Big]^{-1}\Bigg[\bar{\nabla}^2+\frac{4\bar{\Lambda}_f}{3}\Bigg]^{-1}T
\\ \nonumber
&+& \frac{1}{9
f''(\bar{R})}\tilde{T}\Bigg[\frac{1+f'(\bar{R})}{3f''(\bar{R})}
-\Big(\bar{\nabla}^2+\frac{4\bar{\Lambda}_f}{3}\Big)\Bigg]^{-1}\Bigg[\bar{\nabla}^2+\frac{4\bar{\Lambda}_f}{3}\Bigg]^{-1}T.
\eeqa We note that as is shown in Eq.(\ref{bcurv}),  the effective
cosmological constant $\bar{\Lambda}_f$ is not an independent
quantity but it  is determined by $f'(\bar{R})$ and $f''(\bar{R})$.

\section{ van DVZ discontinuity}
The expression of (\ref{opamp}) is quite a nontrivial integral, but
we can study the particle spectrum of graviton and scalar in $f(R)$
gravities  by investigating  the pole structure of the amplitude.
 We note that in $f(R)$ gravity, taking
a limit of $\bar{\Lambda}_f \to 0$  is equivalent to the limit of
$f(R)\to 0$, which is nothing but GR.  Here we read off three poles
from (\ref{opamp}). We wish  to compute  the residue at each pole.

(a) Pole at $\bar{\nabla}^2=-\frac{4\bar{\Lambda}_f}{3}$

The residue at this unphysical pole is zero as \beq
-\frac{2\bar{\Lambda}_f}{9(1+f'(\bar{R}))}
\frac{1}{\Big[-\fr{4\bar{\Lambda}_f}{3}+2
\bar{\Lambda}_f\Big]}+\frac{1}{3(1+f'(\bar{R}))}=0. \eeq

(b) Pole at $\bar{\nabla}^2=-2\bar{\Lambda}_f$

The residue at this physical pole takes the form \beq
\frac{1}{1+f'(\bar{R})}
\Bigg[\frac{2}{3}-\frac{2\bar{\Lambda}_f}{9}\Big(-\frac{3}{2\bar{\Lambda}_f}\Big)\Bigg]=\frac{1}{1+f'(\bar{R})}.
\eeq We emphasize that the residue is positive  only for
$1+f'(\bar{R})>0$, indicating the case that the ghost is absent. In
the limit of $\bar{\Lambda}_f \to 0(f(R)\to 0)$, the residue is $1$,
and thus, the amplitude describes a massless graviton with 2 DOF
like \beq \label{massl-amp}\lim_{\bar{\Lambda}_f\to 0}
\Big[4A\Big]=2\Bigg[
\tilde{T}_{\mn}\frac{1}{-\bar{\nabla}^2}T^{\mn}-\frac{1}{2}
\tilde{T}\frac{1}{-\bar{\nabla}^2}T\Bigg]. \eeq

(c) Pole at
$\bar{\nabla}^2=\frac{1+f'(\bar{R})}{3f''(\bar{R})}-\frac{4\bar{\Lambda}_f}{3}\equiv
m^2_f$

We have a newly massive scalar propagation unless $f'(\bar{R})=0$
and $f''(\bar{R})=0$, which shows the equivalence between $f(R)$
gravity and scalar-tensor theory on AdS$_4$ space.  This pole was
first pointed in Ref.\cite{NS}. Actually, the presence of this pole
reflects that we are working with $f(R)$ gravities.  This pole never
appears  in GR with cosmological constant. A similar pole appears
also when including  $\alpha R^2$~\cite{Neu,GT}.   In this case, we
may regard $\alpha R^2$ as one of $f(R)$ forms.  The residue at this
massive physical pole takes the form \beq
\frac{1}{3(1+f'(\bar{R}))}. \eeq We note that this residue is
positive definite only for $1+f'(\bar{R})>0$, showing the ghost-free
pole. In the limit of $\bar{\Lambda}_f \to 0(f(R)\to 0)$, the
residue is $1/3$, and thus, the amplitude reduces to  \beq
\lim_{\bar{\Lambda}_f\to 0} \Big[4A\Big]=2\Bigg[
\tilde{T}_{\mu\nu}\frac{1}{-\bar{\nabla}^2}T^{\mu\nu}-\frac{1}{3}
\tilde{T}\frac{1}{-\bar{\nabla}^2}T\Bigg], \eeq which shows the van
DVZ discontinuity clearly when comparing to (\ref{massl-amp}).

 As was
mentioned in Ref.\cite{Neu}, our starting action (\ref{act-0}) with
 external sources  provides a massless graviton with 2 DOF for
$f(R)=0$ on AdS$_4$ space, while it provides a massless graviton
with  a massive scalar with 3(=2+1) DOF when choosing $f(R)=\alpha
R^2$ on AdS$_4$ space.

 In this
work, we have shown that  a massless  graviton $h^{TT}_{\mn}$
(5$\to$ 2 DOF, after imposing the source-conservation law
appropriately) and a massive scalar $\psi$ (1 DOF) propagate on
AdS$_4$ space using arbitrary $f(R)$ gravity model with external
sources. Consequently, we show that there is the vDVZ discontinuity
in $f(R)$ gravity models because $R+f(R)$ means GR with an
additional scalar. In the limit of $\bar{\Lambda}_f \to 0(f(R)\to
0)$, we did not recover  the one-particle amplitude for a massless
graviton, but we  did recover the massless limit of one-particle
amplitude for a massive graviton. Also, there is no apparent absence
of the discontinuity since we did not introduce the Fierz-Pauli mass
term on AdS$_4$ space.

In the constant curvature background of $f(R)$ gravities, the
combination of $f(\bar{R})$ and $f'(\bar{R})$ determines the
effective cosmological constant $\bar{\Lambda}_f$, while the
combination of $f(\bar{R})$, $f'(\bar{R})$ and $f''(\bar{R})$
determines the mass squared $m^2_f$ of an  additional scalar in the
perturbation analysis. Hence,  we have recovered general relativity.

Finally, we wish to mention that there was also  physical
non-equivalence between GR and $f(R)$ gravities on different
considerations and purely classical arguments. This has been
observed in cosmological viability of $f(R)$ gravity as an ideal
fluid and its compatibility with a matter dominated
phase~\cite{CNOT}.

\section*{Acknowledgment}
The author thanks Edwin J. Son and Tae Hoon Moon for helpful
discussions.   This work was supported by the National Research
Foundation of Korea (NRF) grant funded by the Korea government
(MEST) (No.2010-0028080).

\end{document}